\documentclass[journal]{IEEEtran}

\usepackage{cite}
\usepackage{amsmath,amssymb,amsfonts}
\usepackage{graphicx}
\usepackage{textcomp}
\usepackage{xcolor}
\usepackage{url}
\usepackage{booktabs}   
\usepackage{array}      
\usepackage{booktabs,array,xcolor,colortbl}
\usepackage{subcaption}
 \usepackage[T1]{fontenc}
 \usepackage{pifont}
 \usepackage{booktabs}
 \usepackage{array}
 \usepackage{xcolor}
 \usepackage{colortbl}

\usepackage{hyperref}
\hypersetup{colorlinks=true, linkcolor=blue, citecolor=blue, urlcolor=blue}

\begin{document}

\title{Heterogeneity-Aware Belief Synchronization for Semantic Communication in AI-Native 6G Networks\\
}

\author{Muhammad Hannan Akram, 
        Muhammad Abubakar Rashid, 
        Wassi Haider Kabir, 
        Haejoon~Jung,~\IEEEmembership{Senior Member,~IEEE},
        Kapal Dev, \IEEEmembership{Senior Member, IEEE}
        and Syed Ali Hassan,~\IEEEmembership{Senior Member,~IEEE}
        
\thanks{Muhammad Hannan Akram, Muhammad Abubakar Rashid, Wassi Haider Kabir, and Syed Ali Hassan are with the School of Electrical Engineering and Computer Science (SEECS), National University of Sciences and Technology (NUST), Islamabad 44000, Pakistan (e-mails: \{makram.bsds23seecs, mrashid.bsds23seecs, wkabir.bee22seecs, ali.hassan\}@seecs.edu.pk).}
\thanks{Haejoon Jung is with the Department of Electronic Engineering, Kyung Hee University, Yongin 17104, South Korea (e-mail: haejoonjung@khu.ac.kr).}
\thanks{K. Dev is with the CONNECT Centre and the Department of Computer Science, Munster Technological University, Ireland (e-mail: kapal.dev@ieee.org)}
}


\maketitle

\begin{abstract}
6G networks will not be serving as communication infrastructures only; rather, they are expected to evolve into intelligent systems, where thousands of autonomous artificial intelligence (AI) agents are interconnected. The agents are deployed across a wide range of platforms including low Earth orbit (LEO) satellites, high-altitude platforms (HAPs), unmanned aerial vehicles (UAVs), edge servers, and terrestrial devices. These agents continuously observe their environment and exchange information. Semantic communication provides an efficient mechanism for exchanging meaningful information instead of raw data. However, its effectiveness depends on the communicating agents having sufficiently aligned beliefs to correctly interpret and decode the transmitted messages. This assumption becomes difficult to satisfy in the 6G network where heterogeneous AI models operate under diverse computational constraints and continuously acquire different knowledge from their local environments. This article presents a heterogeneity-aware belief synchronization framework for 6G AI-native networks. It uses latent translation models deployed on multi-access edge computing (MEC) servers. These models translate belief updates from one agent to agent-specific knowledge without requiring joint training and a homogeneous architecture of models. By exchanging compact belief updates through a latent translation model only when necessary, the framework preserves privacy, reduces synchronization cost, and minimizes local knowledge drift. We validate the framework through a case study on a multi-layered terrestrial/non-terrestrial network. Results demonstrate that it maintains low synchronization cost, measured by the number of parameters transmitted, and low belief alignment error across the heterogeneous agents in the case study. 
\end{abstract}



\section{Introduction}
Sixth-generation (6G) networks are being designed around a fundamentally different paradigm. In addition to making communication faster, 6G networks are expected to interconnect a vast amount of autonomous and AI-powered devices. 
These devices include drones, factory robots, self-driving vehicles, environmental sensors, and stratospheric HAPs. Another feature of 6G is the integration of non-terrestrial networks (NTNs) with traditional network systems \cite{deng2026ntncoor}. Satellites, HAPs,  UAVs, and terrestrial base stations work together as a unified network spanning space, air, and ground. Each of these platforms will not simply send and receive data, but it will host its own AI agent, capable of perceiving, reasoning, and acting according to its understanding \cite{letaief2019}. This shift highlights the concept of AI-native networks, in which AI is no longer treated as an external capability but is integrated within the network. 

The integrated 6G network spans resource-constrained IoT sensors and UAVs to HAPs, satellites, and edge servers. These platforms differ significantly in their computational resources, energy consumption, memory, and communication capabilities. AI agents are selected according to the computational capabilities and application objectives of their hosting platforms.
Low Earth orbit (LEO) satellites operate under strict payload, power, and energy constraints. These constrain them to lightweight AI models for tasks such as sensing, event detection, and simple decision-making. At the \textit{stratospheric  layer}, higher computational capacity allows the use of LLMs capable of complex reasoning and planning. The \textit{terrestrial layer} is the most heterogeneous layer comprising devices that range from resource-constrained devices running lightweight AI models to servers capable of running LLMs \cite{li2025secfft}. These diverse computational constraints and the nature of tasks across the network lead to the deployment of heterogeneous AI agents.

\definecolor{headerblue}{RGB}{219,229,238}
\definecolor{rowgray}{RGB}{245,246,248}
\definecolor{cellgreen}{RGB}{220,245,220}
\definecolor{cellred}{RGB}{250,220,220}
\definecolor{cellyellow}{RGB}{255,245,210}
\definecolor{bordercolor}{RGB}{180,180,180}

\setlength{\arrayrulewidth}{0.6pt}

\begin{table*}[t]
\centering
\renewcommand{\arraystretch}{1.65}
\setlength{\tabcolsep}{6pt}
\begin{tabular}{%
  !{\color{bordercolor}\vrule}>{\raggedright\arraybackslash}p{2.4cm}
  !{\color{bordercolor}\vrule}>{\centering\arraybackslash}p{2.8cm}
  !{\color{bordercolor}\vrule}>{\centering\arraybackslash}p{2.8cm}
  !{\color{bordercolor}\vrule}>{\centering\arraybackslash}p{2.8cm}
  !{\color{bordercolor}\vrule}>{\centering\arraybackslash}p{2.8cm}
  !{\color{bordercolor}\vrule}%
}
\arrayrulecolor{bordercolor}
\hline
\rowcolor{headerblue}
\textbf{Approach} &
\textbf{Heterogeneity} \newline
{\footnotesize Supports different model architectures?} &
\textbf{Knowledge Drift} \newline
{\footnotesize Sync occurs only when belief is changed?} &
\textbf{Sync Cost} \newline
{\footnotesize Overhead less than full parameter sharing?} &
\textbf{Privacy} \newline
{\footnotesize No raw data or full weights exposed?} \\
\hline

DeepSC~\cite{xie2021deepsc}
  & \cellcolor{cellred} Requires identical architecture trained on same dataset
  & \cellcolor{cellgreen} No synchronization mechanism after deployment
  & \cellcolor{cellred} Full model shared during joint training
  & \cellcolor{cellgreen} No exchange after deployment \\
\hline

\rowcolor{rowgray}
TCLSC~\cite{wang2025tclsc}
  & \cellcolor{cellred} Authors explicitly require ``same network structure'' for both sender and receiver
  & \cellcolor{cellred} Parameters shared on a fixed periodic schedule only
  & \cellcolor{cellgreen} Quantized weights are shared
  & \cellcolor{cellgreen} Only quantized weights shared; raw data not transmitted \\
\hline

SKBS~\cite{lu2024skbs}
  & \cellcolor{cellyellow} Supports different model sizes but requires a common global KB schema
  & \cellcolor{cellred} Server randomly selects devices each round; syncs even when KB is unchanged
  & \cellcolor{cellgreen} Compressed global KB shared instead of full models
  & \cellcolor{cellyellow} Network parameters shared with server; raw data not shared \\
\hline

\rowcolor{rowgray}
HeteroKB~\cite{zhang2025heterogeneous}
  & \cellcolor{cellgreen} Handles heterogeneous KB content 
  & \cellcolor{cellred} Feedback sent every round regardless of actual belief change
  & \cellcolor{cellgreen} Only task performance feedback sent, not full model weights
  & \cellcolor{cellred} Receiver must expose inference outputs to transmitter each round \\
\hline

\rowcolor{headerblue}
\textbf{Proposed}
  & \cellcolor{cellgreen} Latent translation models translate belief updates across different agent types
  & \cellcolor{cellgreen} Sync occurs when change detected; unrelated agents excluded from sync
  & \cellcolor{cellgreen} Compact belief updates sent to relevant connected agents
  & \cellcolor{cellgreen} Model parameters never exposed, share only unrestricted updates \\
\hline

\end{tabular}
\arrayrulecolor{black}
\caption{Comparison of existing belief synchronization approaches.
The measurable criterion for each column is stated in the header.
\colorbox{cellgreen}{\strut~Green~} = criterion met;\quad
\colorbox{cellyellow}{\strut~Yellow~} = partially met;\quad
\colorbox{cellred}{\strut~Red~} = not met.
Cell text states reason from the respective paper.}
\label{tab:comparison}
\end{table*}
\begin{figure*}
    \centering
    \includegraphics[width=1\linewidth]{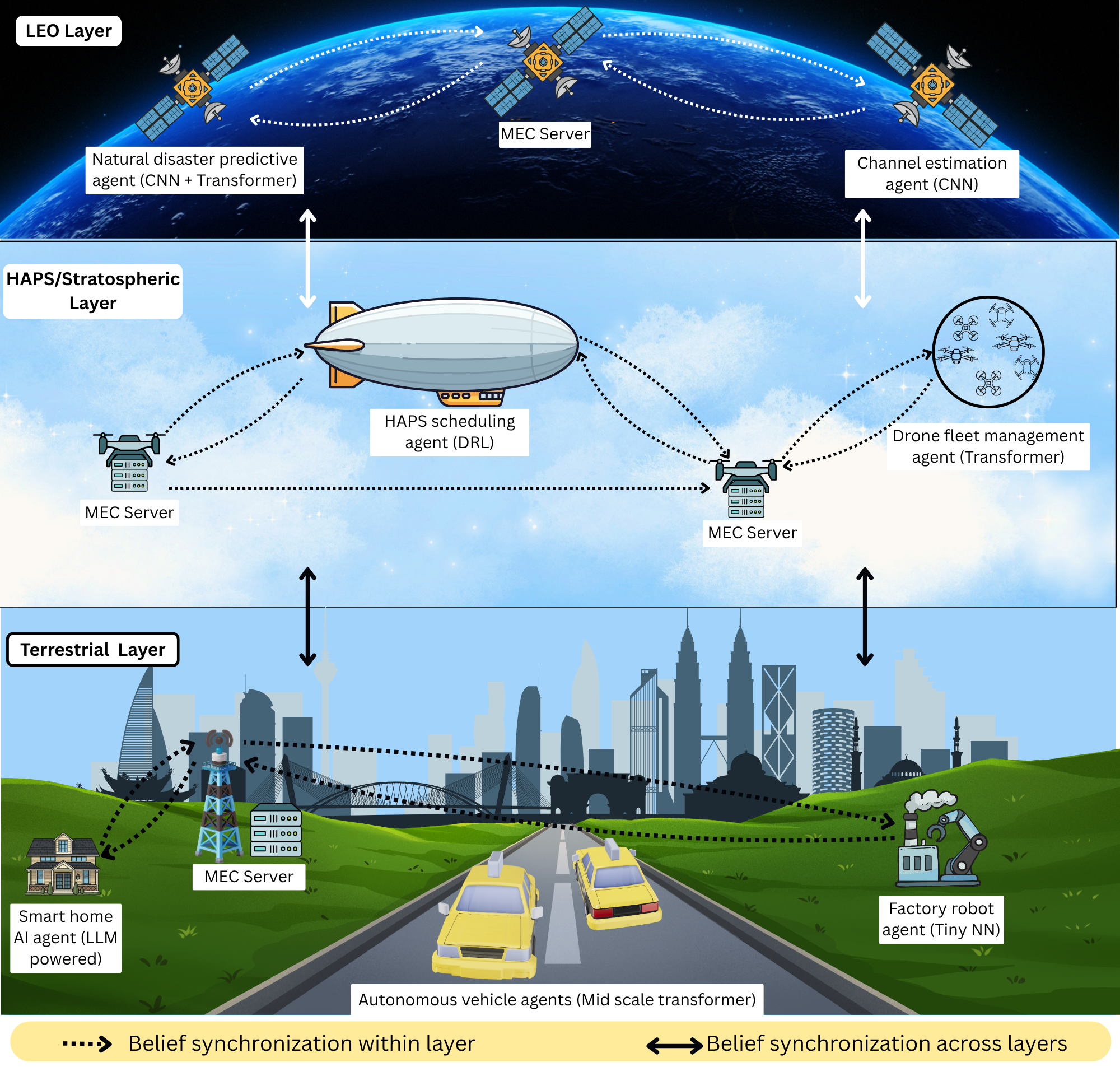}
    \caption{Multi-layer 6G network architecture with heterogeneous AI agents and proposed belief synchronization framework.}
    \label{fig:placeholder}
\end{figure*}
Because these agents differ in architecture, they also develop different internal representations of knowledge. Their beliefs are encoded in distinct latent spaces. Each agent continuously learns from its local environment. A satellite might learn the characteristics of the large regions it observes from hundreds of kilometers above the Earth; a HAP learns from data collected at the stratospheric level. A ground sensor learns only what happens in its immediate surroundings. This local learning is precisely what makes each agent effective at its own task, forming different beliefs. In a network of thousands of learning agents spanning satellites, HAPs, and 
ground devices, no two agents can be assumed to share the same belief of the world they store. This divergence of beliefs across the agents would not be problematic if agents operated independently.
Despite these differences, the agents must exchange information for different applications. A satellite may detect flooding over a large region, a HAP may monitor the condition of roads and bridges within the affected area, while ground sensors measure local water levels and rainfall. Individually, each agent possesses only a partial view of the environment. By sharing their observations, agents can complement one another's knowledge. This allows them to make more informed and coordinated decisions.


Effective semantic communication requires the sender and receiver to maintain sufficiently similar beliefs so that the semantic messages can be interpreted accurately \cite{getu2024survey}. This requirement is particularly challenging in large-scale 6G networks. Such networks integrate terrestrial, aerial, and satellite platforms into a unified AI-native infrastructure. Consequently, AI agents run on devices with widely varying computational resources, model architectures, and operational roles, which makes semantic communication significantly more challenging \cite{yang2023semcomfuture}. Addressing this challenge requires a mechanism that continuously aligns the beliefs of heterogeneous agents so that semantic communication remains reliable over time. However, designing a belief synchronization mechanism requires some interrelated constraints.
\begin{itemize}
\item \textbf{Synchronization cost:} Synchronizing large models (e.g., parameter sharing) can consume significant bandwidth, reducing the resource efficiency of semantic communication.

\item \textbf{Privacy:} Agents may learn from sensitive local data, requiring synchronization mechanisms that share only task-relevant beliefs while preserving privacy.

\item \textbf{Knowledge drift:} Continuous synchronization can weaken an agent's local knowledge, which makes it essential to preserve local beliefs during updates.

\item \textbf{Heterogeneity:} Agents differ in architectures, capabilities, and resource constraints. This requires synchronization methods that operate across diverse models and devices.
\end{itemize}

\begin{figure*}
    \centering
    \includegraphics[width=1\linewidth]{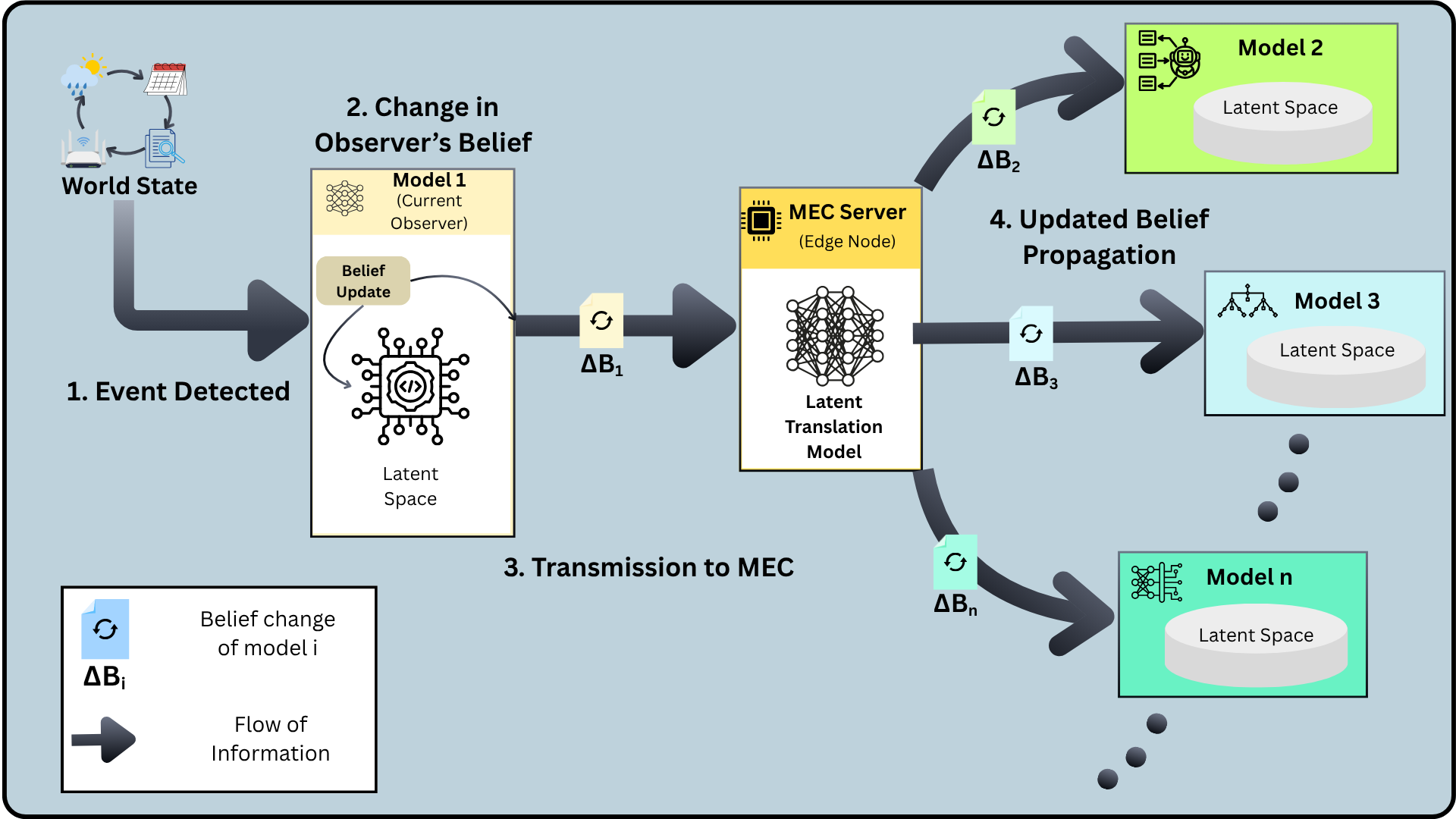}
    \caption{The proposed four-step belief synchronization workflow.}
    \label{fig:framework}
\end{figure*}

Existing studies have addressed some of these challenges individually, as shown in Table \ref{tab:comparison}. However, a unified solution that can handle heterogeneity, privacy, synchronization cost, and knowledge drift together is still missing. To address this gap, this article presents a latent translation-based framework deployed on MEC servers, enabling heterogeneous AI agents to synchronize their beliefs efficiently while supporting reliable semantic communication.



\section{Existing Technologies And Opportunities}\label{existing}
In 6G networks, the exchange of information is expected to be achieved through semantic communication.
In traditional communication, the goal is to accurately transmit the bits to the receiver. However, semantic communication moves beyond this conventional approach \cite{gunduz2023beyond}. 
According to Weaver, communication has three levels: technical, semantic, and effectiveness \cite{shannon1949}. The technical level focuses on the reliable transmission of bits over the communication channel; the semantic and effectiveness levels concern preserving the intended meaning and supporting successful task completion, respectively \cite{guo2024semcomnet}. Semantic communication targets the semantic and effectiveness levels by transmitting only the intended information. Instead of following a "transmit-first-and-understand-later" paradigm, it uses an ``understand-first-and-then-transmit" approach. This extracts and transmits only semantically relevant information while eliminating redundant data.
This represents a shift from "how to transmit" to ``what to transmit," making communication much more efficient and precisely what an AI-native 6G network needs \cite{getu2024goaloriented}, \cite{strinati2021beyond}. However, semantic communication only works under one assumption that the receiving agent can correctly decode and interpret the compressed meaning it receives. For this to happen, the receiving agent must share a significant body of background beliefs with the sending agent to correctly decode and interpret the message. In the semantic communication literature, this belief of an agent about the world is typically referred to as a knowledge base. The sender uses its knowledge base to encode the semantic message, while the receiver uses its own knowledge base to decode and reconstruct the intended meaning.

Early semantic communication systems addressed the belief synchronization requirement by jointly training the sender and receiver models having identical architectures and dataset~\cite{xie2021deepsc}. As a result, agents developed a common internal latent representation. This allowed semantic messages to be encoded and decoded reliably. Within such a controlled environment, semantic communication achieves impressive performance.

\subsection{Belief Synchronization Approaches}

Recent studies have developed strategies for synchronizing the knowledge bases. One of the approaches is parameter sharing, where communicating agents periodically exchange the parameters of their semantic encoder and decoder models over the network. The Transceiver Cooperative Learning (TCL-SC) framework~\cite{wang2025tclsc} follows this strategy by allowing the sender and receiver to cooperatively update their models through parameter exchange. However, this approach assumes that communicating agents share the same neural network architecture. 
Another methodology~\cite{lu2024skbs} addresses belief synchronization using federated knowledge distillation. Instead of exchanging model parameters, each agent shares the knowledge learned by the model to synchronize a global knowledge base. This approach reduces communication overhead, and agents maintain a consistent semantic understanding. However, this approach assumes one global knowledge base is sufficient to encode and decode semantic messages among all the agents in the network. 

\subsection{Research Gap and Opportunities}
These existing studies either use joint training of AI agents or assume homogeneous architecture. In a 6G network, where architecturally different agents are deployed at different devices, joint training and knowledge sharing become infeasible. Moreover, a single global model might not be able to translate semantic messages among thousands of heterogeneous agents.
These limitations highlight the requirement of a mechanism that allows seamless belief synchronization among the heterogeneous agents without jointly training them. Such a mechanism allows smooth integration of new AI agents without requiring modifications in existing AI agents. We propose such a mechanism next.

\section{Heterogeneity-Aware Belief Synchronization Framework}\label{framework}
The proposed framework enables belief synchronization among heterogeneous agents
by introducing latent translation models. These models translate the belief updates of the sending agent into agent-specific belief updates for each receiving agent. 
These latent translation models are deployed on MEC servers distributed across the network.
By using MEC servers, the computationally intensive semantic translation task is offloaded from the network devices so they continue to remain efficient.

\subsection{Belief Synchronization Workflow}
The synchronization process consists of four steps as shown in Fig.~\ref{fig:framework}.

\begin{enumerate}
    \item \textbf{Observation:} Each AI agent continuously observes its local environment and updates its internal beliefs based on its new belief about the world.

    \item \textbf{Belief Update Generation:} Rather than transmitting entire model parameters, the agent transmits a belief update, denoted by $\Delta B$.

    \item \textbf{Semantic Translation:} The belief update is transmitted to its local MEC server, where the latent translation model converts it into agent-specific belief updates compatible with the latent representations of the receiving agents.

    \item \textbf{Belief Propagation:} The translated belief updates are distributed only to the relevant agents connected to the MEC server, except the sender. The relevant receiving agents update their belief state upon receiving a belief update.
\begin{figure}
    \centering
    \includegraphics[width=1\linewidth]{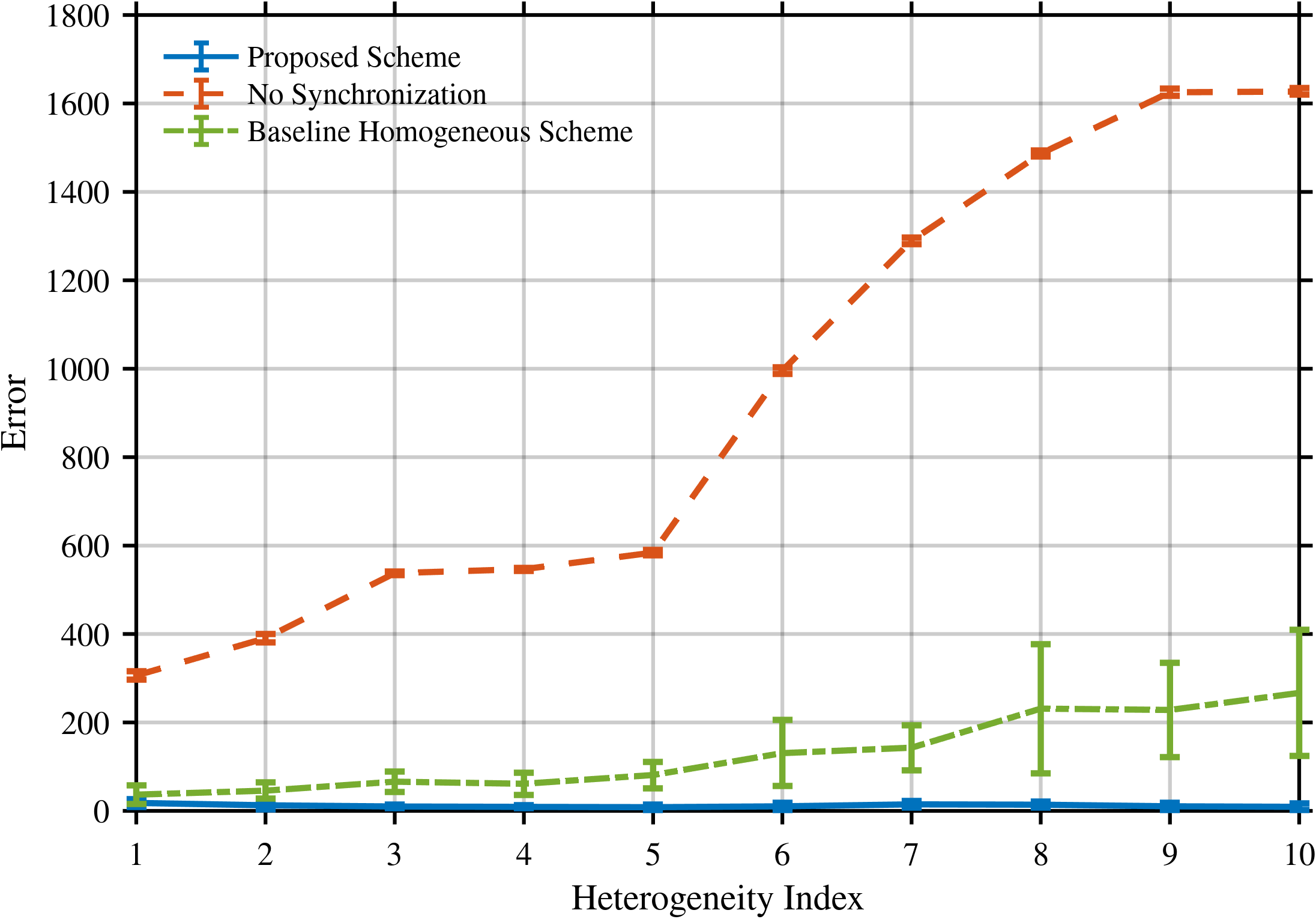}
    \caption{System error trend with increasing heterogeneity}
    \label{fig:fig0}
\end{figure}
\end{enumerate}
In our proposed methodology, heterogeneity is addressed through latent translation models that specialize in translating belief updates from one agent to agent-specific belief updates. This allows small neural networks, LLMs, and other AI agents to synchronize their beliefs. Each MEC server propagates translated compact belief updates only to the relevant connected agents for which the information is important. For example, if a roadside sensor generates a belief update indicating a traffic accident, the MEC forwards this update to nearby autonomous vehicles and traffic management agents for route optimization, while unrelated agricultural drones or environmental sensors connected to that MEC server are excluded.\\ 

Additionally, the belief updates are propagated only when one agent observes a change in its belief. This prevents unnecessary updates, as in the round-based synchronization approaches where the updates are propagated even if there is no change in belief. Such unnecessary updates might cause the receiver to drift from its local knowledge. Exchanging unchecked knowledge or full model parameters might expose private local information. Unlike some existing approaches that exchange complete model parameters, the proposed framework transmits only compact belief updates, thereby reducing raw knowledge exposure.
Furthermore, synchronization is done through MEC servers instead of jointly training agents of the network. This allows natural expansion of the network, as AI agents can be integrated without jointly training with existing agents. During this integration, a small adapter in the latent translation model is trained using calibration examples to align its latent space with the existing representation space. After training, the agent can exchange translated belief updates with other agents through the MEC server. This allows new agents to gradually build their beliefs without retraining, making the network more robust over time.  
We next evaluate this framework in a multi-layer 6G scenario.

\begin{figure}
    \centering
    \includegraphics[width=1\linewidth]{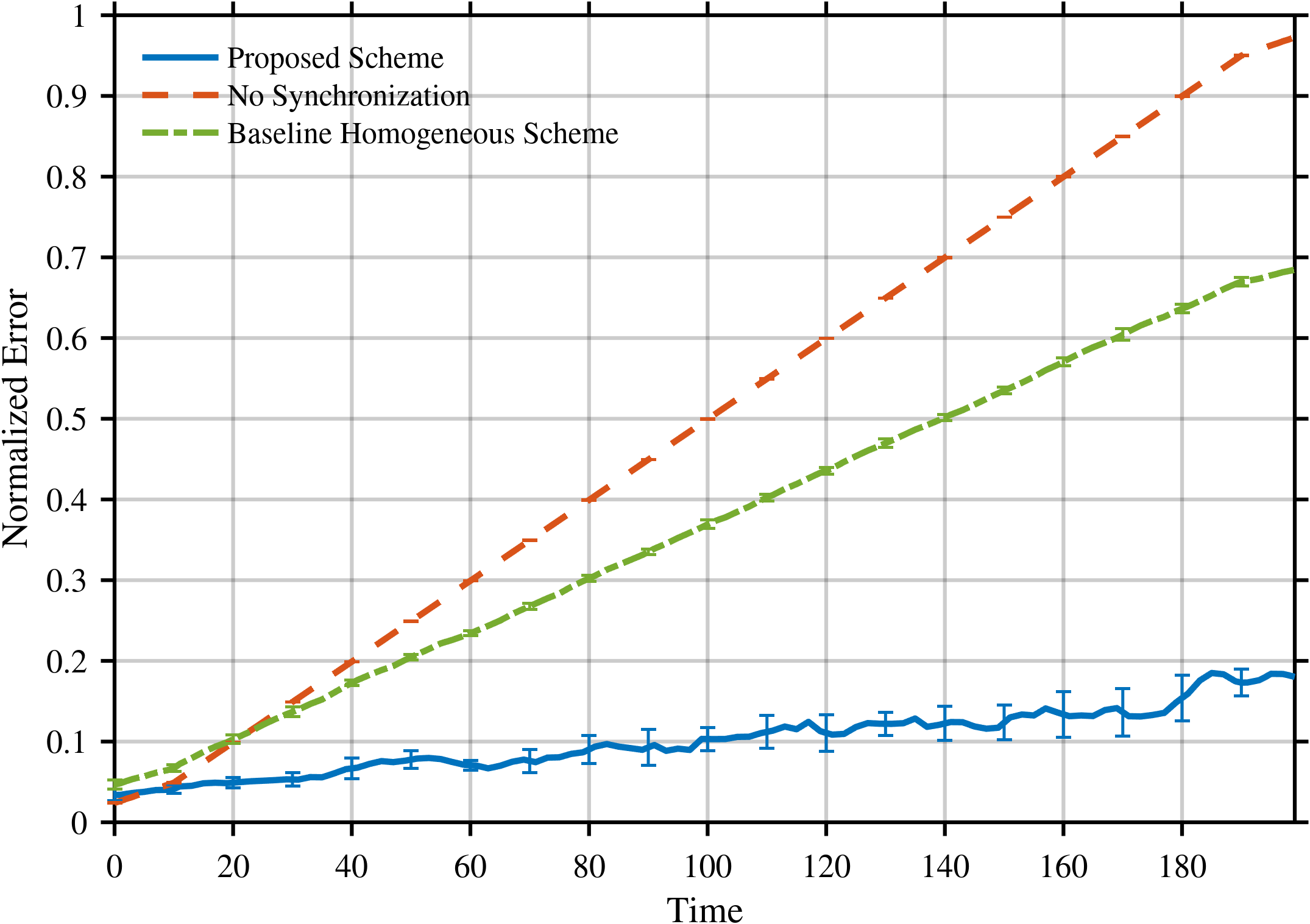}
    \caption{System error trend over time}
    \label{fig:fig1}
\end{figure}

\section{Heterogeneous Belief Synchronization: A Case Study}\label{case}
To validate and analyze the performance of the proposed approach in practice, a 6G TN-NTN communication scenario is considered. In this case study, we consider a multi-tier network consisting of the LEO layer, HAPs/stratospheric layer, and terrestrial layer to model a realistic 6G network scenario. To capture the heterogeneous nature of AI-enabled networks, we assume different agents of varying architectures are present on different devices in all the network layers. All agents are isolated and operate independently. As a result of this heterogeneity and distributed nature, the difficulty in keeping the network synchronized and propagating beliefs becomes much more pronounced. MEC servers hosting the latent translation models are distributed throughout the network.

In our case study, we approximate the world state by a fixed $n$-dimensional vector $\boldsymbol{w}$. An agent is represented as an $n \times m$ matrix where $m$ is the dimension of the agent's latent space. The heterogeneity of the network model is determined by the number of distinct values of $m$ being used by different entities in the system. For instance, a heterogeneity index of 1 means that all the agents in the system are of the same type, whereas a heterogeneity index of 10 means that there are 10 different model architectures in the system, each with a different dimension of latent space. At each time step, $\boldsymbol{w}$ is changed by a fixed magnitude $\delta$ to simulate the changing world state. This change in world state is observed by a model, and the latent space of that model is updated. Since the value of $\boldsymbol{w}$ is known in the system, the latent translation model to convert belief from the latent space of agent $x$ to the latent space of model $y$ is approximated via the matrix $\boldsymbol{X}^{-1}_{n \times m}$. Since $n \neq m$ and the exact inverse of such a matrix does not exist, we use the Moore-Penrose inverse of $\boldsymbol{X}$ to approximate the latent translation model. The change in belief $\Delta{B}$ from agent $x$ to $y$ is obtained via $\boldsymbol{Y}\boldsymbol{X}^{-1}\boldsymbol{x}$ where $\boldsymbol{x}$ is the latent vector of model $x$. The latent space of model $y$ is then updated with the change in belief.

We evaluate the performance of our proposed approach by comparing it with different baselines, such as a network without synchronization and homogeneous synchronization. In a network without synchronization, there is no propagation of belief states, and the network operates under the assumption that the world state remains stationary. In the baseline homogeneous scheme, belief synchronization is performed between models of matching architectures only. In both the proposed approach and the baseline homogeneous scheme, the updates have the same compute budgets and are represented as $\Delta{B}$.

The simulations consider a world state where $n = 1024$ and each heterogeneous agent is assigned a different value of $m$ to reflect the difference in model architectures. World state updates are initiated every second timestep, and at each update event, one agent is selected uniformly at random to observe, process, and propagate the belief update. To evaluate performance, Monte Carlo simulations are conducted over 10 independent runs, with each run using a different random seed to account for stochastic variations. The error is defined as the difference between the readout obtained by following a given approach and the readout obtained in an ideal case where all models have perfect knowledge of the environment. More specifically, the error is $\lvert \boldsymbol{r}\boldsymbol{x} - \boldsymbol{r}\boldsymbol{\hat{x}} \rvert$ where $\boldsymbol{r}$ is the fixed readout defined for agent $x$, $\boldsymbol{x}$ is the latent vector, and $\boldsymbol{\hat{x}}$ is the latent vector of agent $x$ obtained by directly observing the world state $\boldsymbol{w}$.
\begin{table}[t]
\centering
\renewcommand{\arraystretch}{1.8}
\setlength{\tabcolsep}{4pt}

\begin{tabular}{%
  >{\raggedright\arraybackslash}p{2.7cm}
  >{\centering\arraybackslash}p{2.5cm}
  >{\centering\arraybackslash}p{2.8cm}
}
\toprule

\rowcolor{headerblue}
\textbf{Approach} &
\textbf{Average Error $\downarrow$} &
\textbf{Total Communication Traffic $\downarrow$} \\

\midrule

Full Model Transfer &
$0.000 \pm 0.001$ &
$1.59 \times 10^{6} \pm 12.7 \times 10^{3}$ \\

\rowcolor{rowgray}
Baseline Homogeneous &
$0.385 \pm 0.004$ &
$\mathbf{1.08 \times 10^{3} \pm 9}$ \\

Proposed Scheme &
$\mathbf{0.123 \pm 0.013}$ &
$\mathbf{1.08 \times 10^{3} \pm 9}$ \\

\bottomrule
\end{tabular}

\caption{Comparison of synchronization approaches in terms of error and communication overhead. Results are reported as mean $\pm$ standard deviation over multiple random seeds. Communication traffic is measured in the number of parameters transmitted and reported as mean parameters $\pm$ standard deviation.}

\label{tab:communication_comparison}

\end{table}

Fig. \ref{fig:fig0} illustrates the error of all agents in the network with increasing heterogeneity in the system. The heterogeneity index denotes the number of different model types and architectures in the system, with a higher index corresponding to a more heterogeneous network. The results show that as the number of different architectures increases, the overall error of the system increases as well for both baseline approaches. We observe that the error for no synchronization experiences a sharp increase as the heterogeneity index crosses 5 and then later plateaus. The uncertainty bars show a greater variation in the baseline homogeneous approach as compared to the other methodologies. In contrast, the proposed approach keeps the error relatively stable even as heterogeneity increases.

Fig. \ref{fig:fig1} presents the normalized average error of all agents in the network as it progresses through time. Here, each time step corresponds to one simulation iteration, during which agents update their local states and estimates. The normalized error is computed by dividing the average estimation error by the maximum error observed across all methods and time steps. We observe that the naive approach of considering a stationary world state leads to a uniform and steep increase in the error, which shows minimal variation as reflected by the narrow uncertainty bars. This reflects the divergence in the world state, and since synchronization is not supported by this approach, the error keeps on increasing at a steady rate. The same trend is observed in the baseline homogeneous scheme, although to a lesser degree. Since models of only the same architectures are being synchronized, this leaves a large part of the network to continue operating with stale beliefs, and only a subset of the network is kept up to date. In contrast, the proposed approach maintains a relatively stable error over time with a minimal rate of increase. However, the uncertainty bars for the proposed approach become wider, indicating that the variation of errors increases as the world state keeps on diverging over time.

    

Belief synchronization is vital to reflect the changing world state throughout the network. However, communication cost must also be kept in balance so that synchronization does not incur excessive overhead. As shown by Table \ref{tab:communication_comparison}, full model offloading achieves the lowest error but consumes excessive bandwidth due to repeated transmission of all model parameters each time there is an observed change in the world state. Both the proposed and homogeneous approaches have comparable communication cost since they avoid complete transmission of model parameters and instead utilize latent representations or delta weights. However, in a heterogeneous setting, homogeneous synchronization is insufficient as it only updates models of similar architecture, as reflected by its large error value. The proposed approach achieves a $68.1\%$ reduction over the baseline homogeneous scheme while maintaining the same communication overhead.

\section{Open Challenges And Future Directions}\label{future}
As the proliferation of AI agents in modern networks continues and the diversity of agent architectures continues to increase, it becomes crucial to adopt an AI-centric perspective on network design. Semantic communication is the first step towards this approach. The proposed semantic communication methodology introduces a new paradigm in which the maintenance of agent beliefs is separated into a dedicated \textit{knowledge layer}. This opens the door to a broad range of opportunities and applications ranging from the development of foundational latent translation models to the design of scheduling protocols that keep the network of AI agents synchronized under real-time bandwidth constraints. In this section, we highlight such challenges and discuss promising directions for future research.
\subsection{Open Challenges in Belief Synchronization}
There are several open challenges that the proposed methodology might face in real-world settings. We present a few such challenges below.
\subsubsection{Belief Conversion Between Latent Spaces}
A fundamental challenge in belief synchronization lies in converting information between heterogeneous latent spaces. Doing so requires the system to align representations across latent spaces of different architectures. Different models learn internal representations of knowledge differently based on their architecture and training objectives. Similar concepts might occupy distinct regions in the embedding spaces of dissimilar models. This lack of alignment makes translation between latent spaces non-trivial and necessitates mechanisms that can effectively understand the misalignment and bridge the gap in understanding between different architectures. This becomes increasingly difficult to do as the complexity and size of models continue to grow.
\subsubsection{Latent Translation Model Architecture}
The crux of the proposed methodology is the latent translation model, responsible for aligning the latent representation between models, thus translating the change in beliefs from one latent space to another. This cannot be performed using simple multilayer perceptrons (MLPs) because MLPs are trained on fixed input/output sizes and have an inherent feed-forward structure. However, in the proposed methodology, the latent translation model must be able to take an input from a model of any dimension and produce translated outputs of different shapes as required by the architectural dimensions used in the network. Thus, existing methodologies do not possess the flexibility to execute this task. This challenge must be approached from an engineering point of view, and a unique architecture capable of performing cross-model translation effectively must be developed.
\subsubsection{Belief Synchronization Scheduling}
The world state continuously changes in real-world settings. While simulations often use discrete time intervals to update states and synchronize knowledge, real-world scenarios require optimal scheduling in continuous time. Decisions on when to probe the world, propagate knowledge, and accumulate updates before synchronization directly affect system performance. Frequent synchronization improves consistency but increases computational and bandwidth costs, creating a tradeoff that must be balanced based on the application requirements.

 \subsubsection{Prevention of Catastrophic Forgetting}
 The proposed methodology entails that each time belief synchronization is initiated, the belief states of all the agents in the system would undergo an update, and accordingly, the weights and biases would change as well. This is the fundamental mechanism by which a model is able to refresh its belief and incorporate the world state in its latent space. However, the process of updating a model's parameters is non-trivial, as carelessly changing them may lead the weights to change very drastically, ultimately reaching a point where the model loses all sense of the task it was initially trained on, and the agent starts producing erroneous outputs. Thus, knowing which parameters to change and which to preserve is an important challenge of ensuring the effectiveness of the proposed approach.
 \subsubsection{Addressing Malicious AI Agents}
 The case study assumes an ideal network with no adversaries. Such an assumption cannot be taken while implementing the framework in real-world scenarios. A malicious party might take advantage of the system's dependence on belief state update and might initiate garbage updates to degrade network performance or meticulously craft updates in order to skew the collective belief of the system to achieve its pernicious goals. 
 A compromised knowledge layer will allow the attacker to gain complete control over the entire AI operations across the networks. Thus, identifying such malicious updates remains a critical challenge in order to ensure the security of the system and prevent belief distortion and corruption in the system

 \subsection{Future Research Directions}
 Belief synchronization introduces a new paradigm in semantic communication. As a result, we identify the following promising research directions.
 \subsubsection{Latent Space Compatibility and Alignment}
 Translating a change in belief from one latent space architecture to another requires a thorough understanding of not only the model's configuration but also the learning process and what exactly is being learned by the model. There is a need for methods that can be used to probe the embedding space of a model and compare latent representations across different model architectures. For instance, transformers distribute learned representations across all trainable components, including attention and feed-forward layers, while convolutional neural networks distribute them across convolutional filters and other trainable layers. Similarly, autoencoders, diffusion models, and other deep learning models represent learned information through architecture-specific parameterizations and intermediate activations. 
 
 Thus, there is a need to investigate whether all architectures learn the same patterns universally and, if not, what is the fundamental difference between their learned belief states. Although some studies have been conducted in recent years, the focus remains mostly on large language models. A more holistic study is required that attempts to understand the geometry of the learned embedding space across different model architectures and types.
 \subsubsection{Efficient And Mathematically Valid Translation Model Architectures}
 The latent translation model proposed in this study remains an open and promising research direction. It is equally important to focus on task performance and model efficiency since knowledge updates must be processed in real time and extended delays may lead to the system having stale knowledge. The issues with existing architectures are further exacerbated when we consider a dynamic topology in which agents join and leave the network randomly. In such a scenario, the latent translation model must also adapt and must avoid performing redundant calculations. Using a simple neural network for each possible pairing of model architectures becomes impossible to manage. Thus, such an architecture needs to be engineered that caters to these scenarios while maintaining computational costs.  
 \subsubsection{Extending Framework By Incorporating ISAC}
 Our current framework only performs communication. It can be extended by integrating sensing as well, thus making it a semantic integrated sensing and communication (ISAC) system. Such an extension would introduce a new dimension of obtaining knowledge apart from explicitly observing the world state. In turn, the additional sensing capability can improve system security by allowing the system to preemptively detect an attacker and terminate contact, thereby preventing it from compromising the belief synchronization process.
 \subsubsection{Joint Optimization of Synchronization and Resource Allocation}
 In a 6G network, belief synchronization becomes a vital part of network functionality. As is the case with other network operations, belief synchronization must also operate under the usual network constraints such as bandwidth, power, and other channel condition constraints. Thus, instead of treating belief synchronization in an isolated manner, we must optimize it jointly with other network operations. This introduces a new dimension to the optimization problem and would lead to new relationships and tradeoffs between action variables and decision spaces of scheduling problems.

 \section{Conclusion}\label{concl}
In this article, we introduced a new paradigm for semantic communication among heterogeneous AI agents in AI-centric 6G networks. We addressed the issue of heterogeneity in a network consisting of models of different complexities and architectures. By proposing the idea of belief synchronization, we introduced a knowledge layer that operates separately from the other layers in the network. Based on the idea of belief synchronization, we proposed a framework to effectively incorporate it into a network. Owing to the computational complexity of latent belief translation, we propose an MEC offloading scheme that helps reduce the burden on edge devices and instead moves the computation to a separate dedicated machine. We validated the proposed methodology by conducting a case study in which we evaluated its performance and compared it to existing methodologies such as homogeneous synchronization. The results demonstrate that the proposed methodology maintained lower error in the simulated setting. The case study was conducted using a simplified linear matrix–vector representation, providing a controlled environment for evaluation. We conclude the study by identifying several future challenges and research directions.







\end{document}